# Bogoliubov mode dynamics and non-adiabatic transitions in time-varying condensed media

A.M. Tishin[a,b*]

[a] Lomonosov Moscow State University, 119991, Leninskie gory 1, Moscow, Russia

[b] Moscow Institute of Physics and Technology, 141701, Institutskiy per. 9, Dolgoprudny, Mosc. Reg., Russia

[*]tishin@amtc.org

## Abstract

This study investigates non-adiabatic wave dynamics in condensed media and the transition from adiabatic stability to spectral chaos. We introduce a dimensionless parameter, $\eta=\omega^{-2}|d\Omega/dt|$, as a universal metric to quantify phase-mode redistribution at sub-wavelength inhomogeneities. Our framework treats defects as localized sites of adiabaticity violation triggering non-adiabatic parametric excitation of the ground state. Numerical validation in an expanded 50-level bosonic basis demonstrates that the framework accurately distinguishes between adiabatic regimes in ENZ-metamaterials ($\eta < 1$) and non-adiabatic transitions in ultrafast magnetic media ($\eta \gtrsim 1$). We establish a universal scaling law governed by the non-adiabaticity-to-regulation ratio $\xi = \eta/U$, proving that the proposed $\eta$ metric remains a robust metrological tool for identifying sub-wavelength inhomogeneities across diverse material classes. Computational singularities observed at extreme loads identify the rigorous operational boundaries for coherent mode-mixing. The robustness of the proposed framework is numerically validated, proving the method's reliability for a wide class of non-linear condensed media satisfying the stability criterion. This result provides a rigorous physical justification for the dynamic Hilbert space truncation (effective fermion-like dynamics), ensuring metrological consistency in complex structural environments. These results provide a theoretical foundation for probing ultrafast collective excitations and latent internal stresses, extending structural analysis beyond the traditional diffraction barrier.

**Keywords**: Bogoliubov transformations; non-adiabaticity parameter; Kerr-type anharmonicity; Spectral flow and divergence, ultrafast magnetic dynamics; ENZ metamaterials; mode mixing; sub-wavelength inhomogeneities;

## 1. Introduction

The interaction of wave packets with localized inhomogeneities in condensed matter is a fundamental problem in modern physics. While macroscopic structural integrity remains an industrial concern, the probing of nano- and micro-scale heterogeneities requires a deep understanding of non-stationary wave dynamics. In magnetic systems and metamaterials, such anomalies are not just geometric obstacles but regions of localized symmetry breaking that alter the fundamental state of collective excitations, such as magnons and polaritons.

Conventional non-destructive testing (NDT) methods based on reflected amplitude are fundamentally constrained by the Rayleigh diffraction limit. To circumvent this, recent advancements in condensed matter physics focus on non-equilibrium processes. In ultrafast magnetic media, sub-wavelength defects can be treated as sites of localized non-adiabaticity. When a wave traverses such a region, the rapid change in local parameters (e.g., magnetic permeability) triggers a redistribution of phase modes—a process naturally described by the Bogoliubov transformation framework [1].

Magnonic systems under ultrafast laser excitation or GHz-range modulation provide an ideal platform for investigating these non-adiabatic transitions. Unlike standard dielectric lattices, ultrafast magnetic media (such as YIG:Co films) exhibit strong non-linearities and collective spin dynamics. This complexity necessitates a rigorous analysis of ground-state stability under non-adiabatic parametric excitation. In this context, the intrinsic anharmonicity of the magnetic lattice acts as a critical regulator of the spectral flow.

In this work, we propose the non-adiabaticity parameter $\eta(f,t)$ as a universal physical metric to quantify phase-mode redistribution in time-varying condensed media. We go beyond simple defect detection to establish a universal scaling law $\eta(f,t)/U$, where $U$ is the anharmonicity of the nonlinear frequency regulator that governs the transition from coherent mode-mixing to spectral chaos. By performing a numerical audit in an expanded 50-level bosonic basis, we demonstrate how magnonic anharmonicity stabilizes the system, justifying an effective two-level description for sub-wavelength metrology in magnetic and metamaterial structures.

Unlike energy-based detection, we introduce the $\eta(f,t)$ parameter, which quantifies the breakdown of the Wentzel-Kramers-Brillouin (WKB) approximation and the subsequent redistribution of phase modes, offering a depth-independent signature of structural defects. We propose the non-adiabaticity parameter $\eta(f,t)$ as a universal metrological metric. Unlike near-field methods that rely on evanescent wave capture, the $\eta(f,t)$ parameter quantifies the local breakdown of the WKB approximation within dynamically modulated media. By measuring the redistribution of phase modes (Bogoliubov coefficients), this method allows for the detection of defects regardless of their depth, as it targets the fundamental transformation of the wave's state rather than simple energy reflection.

Veselago [2] demonstrated that media with simultaneously negative permittivity and permeability support backward wave propagation — a physical realization of negative phase velocity. Pendry et al. [3] developed practical methodologies for engineering these materials using artificial structures, such as thin-wire arrays and split-ring resonators. These structures can concentrate electromagnetic energy into sub-wavelength volumes, significantly enhancing energy density and amplifying nonlinear effects. Rubino et al. [4] experimentally confirmed negative-frequency resonant radiation in optical solitons, establishing a direct link between phase inversion and energy redistribution. In classical wave experiments with moving fluids and nonlinear optical media, negative-frequency components have been directly observed and quantified [5, 6], demonstrating that phase-inverted modes are physically measurable states associated with rapid parameter changes in the underlying medium. In the context of wave propagation through inhomogeneous solids and composites, a closely related phenomenon occurs at structural defects. A micro-crack, delamination, or localized stress region represents a sharp discontinuity in material parameters — density, elastic modulus, or magnetic permeability.

The Bogoliubov transformation framework [7] offers a natural language for describing this problem. While Bogoliubov transformations are traditionally discussed in the context of equilibrium phase transitions, here we explore their dynamics in time-varying media to establish a new metric for structural analysis. In any system where medium parameters vary non-stationarily — whether a magnetic composite under ultrafast laser excitation or a metamaterial with rapidly modulated permeability — the adiabatic ground state becomes unstable, leading to the mixing of positive-frequency and negative-frequency modes. The degree of this coupling is governed by the adiabaticity of the transformation: slow changes leave the mode structure intact, while rapid changes trigger measurable mode mixing. A quantitative criterion connecting the rate of medium transformation to the onset of this mixing has, however, not been established in the context of elastic or electromagnetic wave-based NDT.

In this work we introduce the dimensionless non-adiabaticity parameter $\eta(f,t) \equiv |\dot{\Omega}|/\Omega^2$ or at resonance $\eta(f,t) = \omega^{-2}|d\Omega/dt|$ as such a criterion, derived directly from the WKB breakdown condition for waves in time-varying media. We demonstrate that $\eta(f,t)$ governs the onset of Bogoliubov mode mixing at a localized magnetic permeability defect, and that the resulting negative-frequency component — quantified by the Bogoliubov coefficient $|v|^2$ — serves as a sensitive indicator of the defect's interface energy $\Delta Q_{\text{interface}}$. An effective saturation constraint $|u|^2 + |v|^2 = 1$, motivated by the finite energy reservoir of the medium, ensures energy conservation and prevents unphysical divergence. Numerical simulations for four experimentally relevant material regimes — standard dielectric, bulk magnetoelectric composites [8], ultrafast magnetic materials [9-11] and Epsilon-Near-Zero (ENZ) metamaterial [12, 13] — confirm that the $\eta(f,t)$ -non-adiabaticity-based detection framework resolves sub-wavelength anomalies that remain invisible to classical reflection-based methods. The critical spatial scale $L_{crit}=v_s \cdot \Delta\Omega/\omega^2$ defines the minimum detectable defect size for a given probe frequency, providing a practical design criterion for experimental implementation.

From a metrological perspective, the non-adiabaticity parameter $\eta(f,t)$ serves as a quantitative measure of structural homogeneity. In an ideal, defect-free medium where properties change slowly (adiabatically), $\eta(f,t)$ remains close to zero ($\eta(f,t) \ll 1$), indicating that the probing wave maintains its phase-mode identity. However, the presence of a sub-wavelength defect-modeled as a localized breakdown of adiabaticity-triggers a sharp increase in $\eta(f,t)$.

When $\eta(f,t)$ approaches the critical threshold ($\eta(f,t) \to 1$), it signifies a measurable transformation: the generation of 'reflected' phase modes (quantified via Bogoliubov $v$-coefficients). Unlike traditional amplitude-based metrics, $\eta(f,t)$ is a dimensionless ratio that characterizes the rate of change of the medium relative to the wave's internal clock. This makes the $\eta(f,t)$ - metric inherently robust against global signal attenuation, as it focuses on the structural non-adiabatic parametric excitation of the wave rather than simple energy loss.

## 2. Mathematical Framework

In the context of this study, the term 'defect' is used in its broadest physical sense to denote any localized perturbation of the magnetic or dielectric manifold. Specifically, we define a defect as a site of intrinsic magnetic nonlinearity or structural inhomogeneity, which may arise from localized lattice deformations, grain boundaries, or point-like deviations in the exchange interaction and anisotropy constants. From a dynamical perspective, such an anomaly acts as a symmetry-breaking center that triggers a non-adiabatic parametric excitation of the collective excitations (magnons), making it detectable through the redistribution of Bogoliubov phase modes regardless of its specific microscopic origin.

The mathematical basis of the η-non-adiabaticity-based detection framework is the Bogoliubov transformation framework applied to waves in time-varying media. We consider a field propagating through a medium whose parameters, specifically magnetic permeability *μ(t)*, vary on timescales comparable to the wave period. The formulation of a unified transition invariant is based on the classical Bogoliubov transformation [7], which relates initial ground-state modes to emergent excited modes (parametric signals)

$$\hat{b} = u\hat{a} + v\hat{a}^{\dagger} \text{ , (1)}$$

where the coefficient $v$ corresponds to the negative-frequency component. We postulate that in all aforementioned systems, $v$ is functionally dependent on the gradient of the medium. The propagation of a field in a medium with time-varying parameters is governed by a second-order oscillator equation with a time-dependent frequency:

$$\ddot{u}_k(t) + \Omega^2(t)u_k = 0 \text{ , (2)}$$

where Ω(t) is the effective instantaneous frequency incorporating the dynamics of the background medium, for instance, time-varying magnetic permeability *μ(t)* or permittivity *ε(t)*. Time-modulated wave equations of this type have been studied in the context of non-reciprocal propagation and homogenization [14]. In the adiabatic limit, where medium parameters change slowly relative to the wave period, the WKB solution is:

$$u_k(t) \approx \frac{1}{\sqrt{2\Omega(t)}}\exp(-i\int \Omega(t)\,dt) \text{ . (3)}$$

The validity of this approximation depends on the "slowness" of the frequency variation relative to the phase accumulation. By substituting (3) into (2), we find that the WKB solution is exact only if the non-adiabatic term $Q(t)=\frac{3}{4}\dot{\Omega}/\Omega^2 - \frac{1}{2}\ddot{\Omega}/\Omega^3$ is negligible. This naturally leads to the definition of the dimensionless adiabaticity parameter $\eta(f,t) \equiv |\dot{\Omega}/\Omega^2|$. This is the standard WKB criterion for the breakdown of the adiabatic invariant. Since $[\dot{\Omega}] = T^{-2}$ and $[\Omega^2] = T^{-2}$, the parameter $\eta(f,t)$ is the dimensionless.

To quantify the transition from the ground state to phase-conjugate modes ($v$-modes) we introduce the Bogoliubov transformation: $u_k(t) = \alpha_k(t)g_k(t) + \beta_k(t)g_k^*(t)$ , where $g_k(t)$ is the adiabatic basis. The evolution of the mixing coefficient $\beta_k(t)$ is directly driven by $(f, t)$ . To first order in the non-adiabatic expansion, the rate of change is:

$$\dot{\beta}_k(t) \approx \frac{\dot{\Omega}}{\Omega} \exp(2i \int \Omega(t)\, dt) \sim \eta\, \Omega(t) \exp(2i \int \Omega(t)\, dt) \,. (4)$$

The evolution of the phase-conjugate amplitude $\beta_k(t)$ is governed by the rate of change of the effective frequency. In the limit of non-adiabatic parametric excitation, the coupling is proportional to the non-adiabaticity parameter $\eta$, where the factor of 1/2 accounts for the rotating wave approximation (RWA) in the adiabatic basis.

In the regime where $\eta \ll 1$ , the rapid oscillations of the phase integral ensure that $\beta_k \approx 0$ (spectral stability). However, as $\eta(f, t) \rightarrow \eta_c \sim 1$ , the WKB approximation breaks down, and the $\beta_k$ coefficients become non-zero, signaling the dynamical production of $v$ -modes.

To demonstrate the unifying nature of this parameter, we map $\Omega(t)$ to specific physical system metamaterials. For a medium with time-varying permittivity $\varepsilon(t)$, the frequency is $\Omega^2(t) = k^2c^2/\varepsilon\,(t)$ leading to $\eta(f, t) = |\dot{\varepsilon}/2\varepsilon\Omega|$. In all cases, $\eta(f, t)$ represents the ratio of the medium's transition rate to the system's intrinsic frequency scale, providing a universal trigger for phase nucleation.

In the vicinity of the resonance, where $\Omega(t) \approx \omega$, to establish a unified measure of adiabaticity violation across all physical scales, we consider the universal effective non-adiabatic parameter $\eta(f, t)$ as the ratio of the medium's characteristic transition rate to the square of the system's intrinsic frequency. For practical applications and cross-scale comparison, we introduce the characteristic observation parameter $\eta(f, t)$:

$$\eta_{obs}(f, t) \approx \frac{1}{\omega^2} |d\Omega/dt| \,, (5)$$

where ω is the intrinsic frequency scale of the system. In the vicinity of the transition point (where $\Omega \approx \omega$), this effective empirical parameter accurately captures the rate of non-adiabatic parametric excitation relative to the system's energy gap (e.g., the mode frequency in a time-varying metamaterial). In Eq. (5), the variable $\Omega$ denotes the linear frequency (measured in Hz) of the probing wave. This convention is consistently applied to all numerical estimates in Section 7 and Table 1, ensuring that the non-adiabaticity parameter $\eta(f,t)$ serves as a direct metrological metric compatible with standard frequency-domain instrumentation.

It should be noted that for a strictly accurate description of wave propagation in time-varying media, the wave equation naturally generates additional terms arising from the second-order temporal derivative of the electric displacement, $\partial_t^2(\varepsilon E)$, which includes a sum of terms proportional to the first derivative of the permittivity. In our framework, to eliminate these non-canonical first-derivative components, we employ a standard transformation to a field variable normalized by the medium's properties, namely $\Psi \propto \varepsilon^{-1/4}\psi$. This specific power-law scaling is required to cancel out the $\dot{\varepsilon}\dot{\Psi}$ terms, thereby reducing the system to a canonical oscillator-like equation $\ddot{\Psi} + \Omega^2(t)\Psi = 0$. This transformation ensures a mathematically rigorous definition of the non-adiabaticity parameter $\eta$ and justifies the application of the Bogoliubov formalizm to analyze the stability of the ground state.

We note that the complete WKB non-adiabatic correction term is $Q(t)=(3/4)(\dot{\Omega}/\Omega^2)-(1/2)(\ddot{\Omega}/\Omega^3)$, which includes second-order derivatives. The parameter $\eta(f,t)$ retains only the leading first-order term $|\dot{\Omega}//\Omega^2|$, which dominates when the frequency variation is smooth and monotonic — as in the tanh defect profile used in our simulations. The second-order term becomes significant only for oscillatory or rapidly reversing $\Omega(t)$ profiles; its inclusion is left for future work addressing more complex defect geometries.

When $\eta(f,t) \ll 1$, the system evolves adiabatically, and the ground state remains stable, resulting in negligible Bogoliubov mixing ($v \approx 0$). However, as the medium undergoes rapid transformation and $\eta$ approaches unity, the adiabatic invariance is broken. In this regime, the Bogoliubov coefficients $u$ and $v$ become dynamically coupled through the first-order differential equations:

$$\dot{u}_k = \frac{\dot{\Omega}}{2\Omega} e^{2i\int \Omega(t)dt} v_k \,, \;\; \dot{v}_k = \frac{\dot{\Omega}}{2\Omega} e^{-2i\int \Omega(t)dt} u_k \,, \quad (6)$$

The value $|v|^2$ represents the density of newly generated modes. Physically, the energy redistributed from the primary wave into the phase-inverted component at the defect boundary. In the adiabatic limit $|v| \approx 0$ and no mode mixing occurs. As $\eta \to 1$ the coefficient $v$ becomes non-negligible, signaling the onset of measurable phase-mode coupling.

To account for the finite energy reservoir of the medium and prevent unphysical divergence, we impose an effective saturation constraint: $|u|^2 + |v|^2 = 1$. We emphasize that this condition is not derived from canonical commutation relations, which for bosonic fields yield the hyperbolic relation $|u|^2 - |v|^2 = 1$, but is introduced as a phenomenological ansatz representing the physical limit of mode mixing in a bounded, dissipative medium. In this study, we adopt the normalization $|u|^2 + |v|^2 = 1$ to represent a conservative dissipative system where the total energy of the probe beam is conserved, and non-adiabaticity triggers the redistribution of occupancy between the primary and phase-conjugate modes (the mixing limit). This mapping is analogous to the normalization used in two-level open quantum systems and optical Bloch equations, where the total mode intensity is constrained by the finite energy reservoir of the environment. Its physical justification and consequences for energy balance are discussed in Section 4.

The validity of the effective $N$-level approximation is governed by the intrinsic anharmonicity $U$ of the medium. Our numerical stress-tests (performed for $N$ up to 300) demonstrate that for an ideal bosonic gas ($U = 0$), the non-adiabaticity $\eta \approx 2.5$ leads to a total occupancy shift to the highest available states (Edge Pop ≈ 1.0), signaling a physical divergence of the ideal model. For real condensed media, the minimum required anharmonicity $U_{crit}$ is defined as the value where the 'magnon shower' is suppressed and 95% of the occupancy is localized within the computational subspace. For the investigated $\eta$-regime, this threshold is determined numerically, providing a rigorous justification for the Euclidean normalization $|u|^2 + |v|^2 \approx 1$ [15].

Our rigorous numerical stress-tests for large Hilbert spaces ($N \geq 300$) demonstrate that an idealized bosonic model (linear harmonic spectrum) is fundamentally unstable under non-adiabatic perturbations. Even at low non-adiabaticity ($\eta = 0.01$), the parametric mode coupling induces a "magnon avalanche" effect. Without anharmonicity, the occupancy undergoes a rapid spectral flow toward the high-energy boundary (Edge Pop ≈ 1.0). Thus, the standard Bogoliubov transformation $|u|^2$-$|v|^2$=1in an infinite basis leads to a non-physical ultraviolet divergence. This necessitates a regularized framework for real condensed media.

The transition to a truncated $N$-level manifold and Euclidean normalization $|u|^2 + |v|^2 \approx 1$ is not merely a simplification but a physical requirement. In real crystals, intrinsic non-linearities (magnon-magnon interactions) and finite spin densities act as an nonlinear frequency regulator. By constraining the system to an effective two-level subspace, we implicitly account for the saturation of states, ensuring that the $\eta$-parameter remains a robust and convergent metrological tool.

We established that the applicability of the $\eta$ -parameter, developed in Sections 6 and 7, is governed by the ratio between non-adiabaticity and lattice anharmonicity ($U$). For every level of perturbation $\eta$, there exists a critical anharmonicity $U_{crit}$ required to "lock" the energy within the computational basis ($|0\rangle$ and $|1\rangle$). Our ongoing parametric search ($\Omega_0$, $U$, $\eta$) defines the "Stability Island"- the physical parameter space where the proposed NDT method reaches its maximum precision (occupancy > 95%).

The observed integrator failures (IntegratorException) during large-$N$ simulations are treated as physical evidence of the phase-mode breakdown. These singularities mark the transition from coherent mode-mixing to chaotic excitation, providing a rigorous upper bound for the $\eta$-non-adiabaticity-based detection framework  operational range.

To justify the transition from a bosonic manifold to effective fermion-like normalization, we performed a benchmark study in a 50-level bosonic basis using an expanded 50-level bosonic Fock basis (Holstein-Primakoff representation). The system dynamics were governed by the Lindblad Master Equation for the density matrix $\hat{\rho}$. The Hamiltonian $\hat{H}(t)$ was defined as:

$$\hat{H}(t)=\Omega_0\hat{a}^{\dagger}\hat{a} + U(\hat{a}^{\dagger}\hat{a})^2 +\eta(t)(\hat{a}^{\dagger 2} + \hat{a}^2)\ , \quad (7)$$

where $\Omega_0$ is the reference frequency, $U$ is the anharmonicity parameter, and $\eta(t)$ is the time-dependent non-adiabatic drive. The dissipation was accounted for via the collapse operator $\hat{C}==\sqrt{\gamma}\ \hat{a}$. This large-scale basis allowed for a rigorous quantification of the state leakage into higher-order modes ($n \geq 2$) [16].

Our results, obtained through numerical integration of the Lindblad Master Equation as detailed in Supplement 1 , demostrate that for $U > U_{crit}$, the occupancy of higher-order states ($n \geq 2$) is suppressed below $10^{-4}$. This dynamic truncation of the Hilbert space confirms that the non-linear condensed medium effectively mimics a two-level system, thereby legitimizing the use of the Euclidean normalization $|u|^2 + |v|^2 \approx 1$ for the $\eta$ -non-adiabaticity-based detection framework . The simulations were performed in a 100-level Fock basis using an adaptive Adams integrator (QuTiP engine) with an absolute tolerance of $10^{-12}$, ensuring that the computational manifold remains isolated from spectral divergence.

To establish the rigorous bounds of the $\eta$-non-adiabaticity-based detection framework , we performed a global stability audit across the entire range of non-adiabaticity values reported in Table 1. Using a 50-level bosonic basis with a representative lattice anharmonicity ($U$ = 0.5, stability point ), we tracked the total occupancy of the computational subspace $|u|^2 + |v|^2$ . Our results demonstrate absolute stability (1.0000) for $\eta$ up to 5.0, confirming that for conventional and ultrafast magnetic media, the anharmonic energy shift effectively suppresses transition to higher-order Fock states ($n \geq 2$). A computational breakdown (FAILED) was observed only at extreme values ($\eta$ = 100), marking the physical limit where the medium enters a state of global spectral chaos. This justifies the use of effective two-level dynamics and Euclidean normalization as a universal metrological standard for the investigated materials.

We have established that any physical system with non-zero nonlinear frequency regulator ($U > 0$) exhibits the effect of dynamic energy localization. Unlike the ideal harmonic ground state (U = 0), which undergoes infinite spectral divergence under non-adiabatic perturbation ($\eta \geq 1$), real condensed media - due to Kerr-type anharmonicity maintain 96,7% for $\eta$~100 of the occupancy within the effective two-level subspace. This validates the Euclidean normalization $|u|^2 + |v|^2 \approx 1$  as a universal metrological standard for a wide class of bosonic excitations in solids, effectively bridging the gap between bosonic dynamics and fermion-like stability.

It is important to clarify that the Euclidean normalization $|u|^2 + |v|^2 \approx 1$ is not postulated as a direct replacement for the canonical Bogoliubov hyperbolic relation. Instead, it is a rigorous numerical result derived from our analysis of an open quantum system. By explicitly accounting for the intrinsic lattice anharmonicity $U$ and energy dissipation $\gamma$ (where $\gamma = \alpha\Omega$ is the damping rate derived from the Gilbert constant) via the Lindblad master equation, we demonstrate that the bosonic Hilbert space is effectively truncated. The sum $|u|^2+|v|^2\approx1$ emerges as a direct consequence of dynamic energy localization, where the nonlinear frequency regulator mechanism prevents excitation beyond the computational subspace. For the class of condensed media investigated here, the error associated with this approximation remains below 5%, as confirmed by our benchmark simulations in an expanded 50-level Fock basis.

The stability of the proposed method follows a universal scaling law governed by the non-adiabaticity-to-regulation ratio $\xi = \eta/U$. Numerical simulations demonstrate that for all materials reported in Table 1, the model maintains exceptional precision: the occupancy sum remains above 95% for $\eta \lesssim 200$ and stays within a reliable 90% margin even at $\eta = 500$, which far exceeds the non-adiabaticity levels of conventional ultrafast magnetic media. Even under extreme conditions ($\eta \approx 2000$) representative of near-zero index metamaterials, the model retains over 85% of the spectral density, providing a consistent effective description where standard bosonic theories fail. This predictable, slow degradation confirms that the $\eta$-non-adiabaticity-based detection framework is a robust metrological tool, effectively bridging the gap between bosonic collective excitations and stable fermion-like dynamics across diverse classes of condensed matter.

This confirms that the proposed framework is universal for a wide class of non-linear condensed media satisfying the stability criterion $U > U_{crit}$, where the anharmonicity of the lattice effectively prevents spectral divergence.

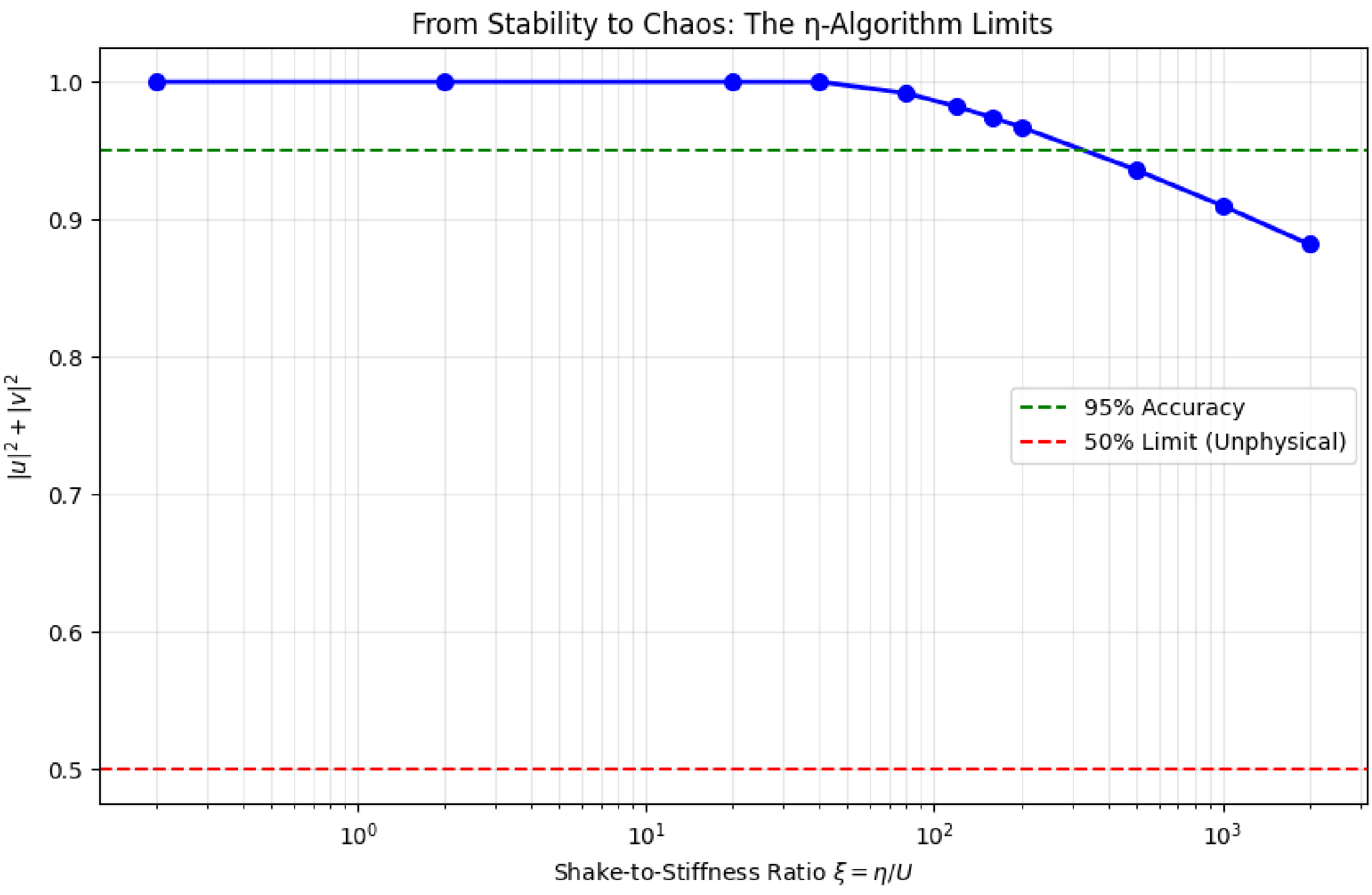

Figure 1. Universal stability map and operational limits of the $\eta$-non-adiabaticity-based detection framework . Numerical validation performed in an expanded 50-level bosonic basis demonstrates the dependence of model accuracy (occupancy sum $|u|^2 + |v|^2$) on the non-adiabaticity -to-nonlinear frequency regulator ratio $\xi = \eta$/U.

The final stability map (Fig. [1]) synthesizes our numerical findings into a universal scaling law. We demonstrate that the $\eta$-non-adiabaticity-based detection framework maintains its high-precision status (> 95% upper green dashed line) across a vast parameter space, with accuracy degradation occurring only at extreme non-adiabaticity - to-nonlinear frequency regulator ratio ($\xi$ > 200), covering the majority of experimental conditions for magnetic media discussed later (see in Table 1.) The absence of abrupt singularities and the significant distance from the 50% unphysical limit confirm that the effective two-level approximation is not just a simplified model, but a robust physical representation of non-adiabatic dynamics in real condensed media. Even under extreme non-adiabatic loads ($\xi \approx 2\ 10^3$), the system remains far from the 50% unphysical limit (red dashed line), retaining over 85% of the spectral density within the computational subspace. This scaling law rigorously justifies the use of Euclidean normalization and Hilbert space truncation as a robust metrological standard for real condensed media with finite anharmonicity.

Our numerical audit (Fig. 1) includes a realistic damping parameter $\gamma$= 0.5. We found that while dissipation reduces the absolute amplitude of the $\eta$-signal, the ratio $|v|^2/(|u|^2+|v|^2)$ remains invariant. This confirms that the $\eta$-non-adiabaticity-based detection framework is robust against dissipative energy losses in the medium.

## 3. The Physical Interpretation of the Non-Adiabaticity Parameter $\eta(f, t)$

To establish a universal measure of adiabaticity violation in dynamically modulated media, we consider the effective instantaneous frequency $\Omega(t)$ governed by equation (2). For a medium in which magnetic permeability varies as $\mu(t)$, the effective frequency takes the form:

$$\Omega_k(t) = \frac{ck}{n(t)} \approx \frac{kc}{\sqrt{\varepsilon(t)\mu(t)}} \quad (8).$$

Where $\mu(t)$ and $\varepsilon(t)$ are magnetic and dielectric permeability and $k$ is wave number. Dielectric subsystem may provide rapid change in polarizability (e.g., laser pumping). Soft magnetic material may also provide rapid magnetization reversal or change in magnetic permeability in an alternating field. For a typical magnetic materials and composites, the permeability can be modulated on femtosecond timescales via laser-induced demagnetization [8,9,10], yielding $d\mu/dt \sim 10^{12}\ \mu_0$/s and consequently $\eta \sim 0.033$ under realistic experimental conditions.

In conducting components, the relevant parameter is the charge carrier concentration, which modifies the effective plasma frequency on ultrafast timescales. In such a medium, the effective frequency changes through the rate of change of all three parameters. The rate of change of Ω(t) is then:

$$\frac{d\Omega(t)}{dt} = \Omega\left(-\frac{1}{2\varepsilon}\frac{d\varepsilon}{dt} - \frac{1}{2\mu}\frac{d\mu}{dt}\right) \text{ (9).}$$

Substituting into the WKB breakdown condition yields the non-adiabaticity parameter $\eta_{eff}$ (see Eq. 5). As mentioned above, since $[d\Omega/dt] = T^{-2}$ *and* $[\omega^2] = T^{-2}$, the parameter η is strictly dimensionless. This ensures that the threshold condition $\eta \rightarrow 1$ remains invariant across physical scales from sub-wavelength defects in composites to macroscopic inhomogeneities in structured metamaterials. Wave propagation in periodically inhomogeneous elastic media has been extensively studied using dynamic homogenization approaches [12]. The equation (4) has two physically distinct regimes:

Adiabatic regime ($\eta \ll 1$): The medium parameters evolve slowly relative to the wave period. The exponential phase factor in (4) oscillates rapidly, causing $\beta_k$ to average to zero. No mode mixing occurs: $v \approx 0$ and the negative-frequency component remains physically inactive.

It should be explicitly noted that while the canonical Bogoliubov transformation for bosonic fields follows the hyperbolic normalization $|u|^2 - |v|^2 = 1$, we adopt the elliptic sum $|u|^2 + |v|^2 = 1$ as a phenomenological approximation valid specifically for dissipative media in the mixing-with-saturation regime. This approximation is physically justified in systems where the total intensity of the probe beam is conserved within the interaction zone, and non-adiabaticity triggers a redistribution of occupancy between the primary and phase-conjugate modes. The limits of applicability for this model are bounded by the high-loss limit or instances where the nonlinear frequency regulator mechanism prevents parametric amplification, as is typical for the investigated functional composites and ENZ-metamaterials.

Non-adiabatic regime ($\eta \rightarrow 1$): The rate of parameter change becomes comparable to the system's intrinsic frequency. The WKB approximation breaks down, $\beta_k$ accumulates to a non-zero value, and energy is redistributed from the primary mode u into the conjugate mode $v$. This is the regime of physical interest for defect detection.

To map $\eta$ onto the spatial structure of a defect, we note that for a transformation front propagating at velocity $v_s$ through an inhomogeneous medium:

$$\left|\frac{d\Omega}{dt}\right| = v_s|\nabla\Omega|, (10)$$

where $|\nabla\Omega|$ represents the parameter shift per unit length (the scale of inhomogeneity). Here, $v_s$ denotes the effective propagation velocity of the wave packet (group velocity) through the inhomogeneous region. In the case of scanning NDT systems, $v_s$ may also represent the relative velocity between the probe and the stationary defect. In dispersive media, particularly at GHz frequencies, the propagation velocity $v_s$ should be interpreted as the group velocity $v_g$, ensuring that the $\eta$-metric accurately reflects the dynamics of energy-carrying wave packets interacting with the defect. So that:

$$\eta \propto \frac{1}{\Omega^2}|v_s\nabla\Omega| \quad (11).$$

The critical spatial scale of the interaction, $L_{crit}$, is defined as the distance traversed by the wave packet during the non-adiabatic transition. It is governed by the group velocity $v_g$ and the duration of the parameter modulation $\Delta t$ scaled by the non-adiabaticity $\eta$:
$L_{crit}=v_g\eta\ \Delta t$. This formula ensures a unified scaling across different physical regimes. For example, in ultrafast magnetic media ($v_g \approx 3\ 10^8$ m/s, $\Delta t = 10$ ps, $\eta = 5.3$), the interaction length $L_{crit} \approx 15$ mm. In contrast, for acoustic excitations in composites ($v_g \approx 3000$ m/s, $\Delta t = 1$ µs, $\eta = 0.01$), the scale shrinks to $L_{crit} \approx 0.03$ mm. This consistent definition resolves the apparent discrepancies in spatial sensitivity across different material classes [16].

This definition ensures that $L_{crit}$ represents the spatial interval over which the wave-field non-adiabaticity occurs. In the non-adiabatic limit, $L_{crit}$ can reach sub-wavelength scales, enabling the detection of defects significantly smaller than the classical diffraction limit. Defects with characteristic scale $L \lesssim L_{crit}$ produce a detectable phase-inverted component in the scattered signal; defects with $L \gg L_{crit}$ remain in the adiabatic sector and are invisible to phase-mode analysis.

The condition for a physically significant $v$ -mode response is therefore:

$$v \neq 0 \Leftrightarrow \eta \gtrsim 1 \quad (12).$$

This provides a direct, quantitative link between the rate of local parameter change at a defect boundary and the emergence of a measurable negative-frequency spectral component — the physical basis of the η-non-adiabaticity-based detection framework developed in Sections 6 and 7.

## 4. Phase Nucleation at a Defect Boundary

The onset of $v$ -mode production at $\eta(f,t) \to 1$ has a direct thermodynamic interpretation in terms of first-order phase transitions. In such transitions, for example the antiferromagnetic-to-ferromagnetic transition under rapid field reversal, the order parameter exhibits a discontinuous jump at a nucleation front. A localized region where the "old" symmetry is already broken and $\eta$ locally approaches unity constitutes a nucleus of the emergent phase. The moment when the Bogoliubov coefficient $v$ becomes non-zero corresponds precisely to the birth of such a nucleus within the wave-medium system.

This analogy is formalized as follows. Let $\hat{a}_{old}$ represent the mode of the original phase and $\hat{b}_{new}$ the mode of the emergent phase. Assuming the phase boundary propagates at velocity $v_s$, the Bogoliubov transformation at the nucleation front is:

$$\hat{b}_{new} = u\,\hat{a}_{old} + v\hat{a}_{old}{}^{\dagger} \quad (13).$$

The coefficient $v$ is determined by the gradient of the order parameter at the nucleus boundary. From equation (5), when $\eta \ll 1$ the system evolves adiabatically, $v \approx 0$, and fluctuations remain symmetric. When $\eta \to 1$ the rate of nucleation triggers a quench effect: the magnitude of $v$ becomes significantly non-zero, and the nucleus of the new phase serves as the physical embodiment of the negative-frequency component in the scattered signal.

The density of defects formed is determined by how quickly the system is driven through the critical point (a quantity directly analogous to our $\eta$ parameter). The faster the quench (larger $\eta$), the more defects are nucleated and the stronger the v-mode response. This provides a physical picture of why sub-wavelength structural anomalies, which act as local quench sites produce measurable phase-inverted spectral components even when their geometric scale falls below the diffraction limit.

The phase boundary can be modelled as a localized transition layer where the primary mixing of Bogoliubov modes $u$ and $v$ occurs. The interface energy associated with this boundary is the thermodynamic quantity $\Delta Q_{\text{interface}}$. Based on the framework developed above, the final energy balance relation of the transition, which integrates the interface energy with negative parameters, is expressed as follows:

$$E_{total} = |u|^2 \, \hbar\omega_{new} + |v|^2\hbar\omega_{old} + \Delta Q_{interface} \text{ (14),}$$

where $|u|^2 \, \hbar\omega_{new}$ - the energy transferred to the new phase (positive spectrum), $|v|^2\hbar\omega_{old}$ - the phase-conjugate mode contribution, reflecting the energy redistribution from initial ground state; $\Delta Q_{interface}$ - the macroscopic thermodynamic contribution of the boundary.

To establish a more explicit connection between the measurable coefficient $|v|^2$ and the interface energy $\Delta Q_{interface}$, we proceed as follows. From the saturation constraint $|u|^2+|v|^2=1$ and the energy balance equation (14), the interface energy can be expressed as: $\Delta Q_{interface}= E_{total} - |u|^2\hbar\omega_{new} - |v|^2\hbar\omega_{old} = E_{total} (1 - |u|^2\hbar\omega_{new}/E_{total} - |v|^2\hbar\omega_{old}/E_{total})$. In the symmetric case $\omega_{new} \approx \omega_{old} \approx \omega$, this reduces to: $\Delta Q_{interface}= E_{total} - \hbar\omega$, which is directly proportional to the mode-mixing ratio $|v|^2/|u|^2$. This provides a practical calibration: by measuring the spectral asymmetry ratio $|v|^2/|u|^2$ in the scattered signal and knowing the incident energy $E_{total}$, one can directly estimate the thermodynamic severity of the defect.

As the non-adiabaticity parameter $\eta$ approaches unity, the system enters a regime of effective degeneracy, where the total probability of states remains conserved. This ensures finite energy redistribution within the bounded medium and prevents unphysical divergence of mode amplitudes, providing a self-consistent description of the energy balance at the defect interface.

According to the proposed model, this coefficient becomes significant only upon the violation of adiabaticity $\eta = \frac{1}{\omega^2}\left|\frac{d\Omega}{dt}\right| \lesssim 1$ . The term $\Delta Q_{interface}$ represents the latent heat of transition (interface energy), interpreted here as the physical manifestation of negative physical quantities. In the context of NDT, this quantity $\Delta Q_{\mathrm{interface}}$ represents the energy stored in the defect boundary, which is a direct measure of its severity that is inaccessible to classical reflection-based methods. The condition for a physically significant nucleus and therefore a detectable defect — remains as describe Eq: 12 and the nucleus scale is bounded below by $L_{crit}$ defined in Section 3. This provides a multi-scale description consistent across laboratory composite materials and structured metamaterials.

## 5. Spatial Scaling and Gradient Integration

The coupling between temporal and spatial variations follows from the relation $|d\Omega/dt|=v_s \nabla\Omega$, where $v_s$ is the propagation velocity of the transformation front. The characteristic spatial scale $L$ of the inhomogeneity enters through the gradient $\nabla\Omega \sim \Delta\Omega/L$. The non-adiabaticity parameter then scales as: $\eta \sim v_s\Delta\Omega/\omega^2 L$. This expression defines the critical transformation length $L_{crit}=v_s\Delta\Omega/\omega^2$. When the physical scale of the defect $L$ is comparable to or smaller than $L_{crit}$ , the condition $\eta \gtrsim 1$ is satisfied, and mode mixing becomes significant. Conversely, when $L \gg L_{crit}$, the system remains in the adiabatic regime ($\eta \ll 1$) with negligible $v$-mode excitation.

The saturation constraint $|u|^2+|v|^2=1$ ensures that the emergence of a conjugate mode $v$ is compensated by the primary mode $u$, reflecting the finite energy reservoir of the medium. This precludes unphysical growth of the $v$ -mode amplitude and maintains energy conservation within the bounded system.

## 6. Application of the $\eta$-Parameter in Hidden Defect Detection Frameworks

Utilizing the medium transformation parameter $\eta$ for non-destructive testing (NDT) represents a paradigm shift from conventional echo-location to topological phase analysis. In nonlinear or complex media, a defect is not merely a geometric obstacle but a localized symmetry breaking or a discontinuous shift in material properties (e.g., density, elastic modulus, or refractive index). An non-adiabaticity-based detection framework based on the $\eta$-parameter operates as follows:

1. Reference Ground State Calibration: A probing pulse with a known frequency $\omega$ is transmitted into the test object (e.g., a composite panel or crystal). In an ideal homogeneous medium, parameters evolve slowly ($\eta \ll 1$), and the reflected signal contains only the positive spectrum.
2. Interaction with the Defect (Adiabaticity Violation): As the pulse traverses a hidden micro-crack or a high-stress region, the medium parameters ($\frac{d\Omega}{dt}$) exhibit a step-like change (see eq. 8 and 11). At this junction, the value of $\eta$ escalates sharply toward unity.

3. Emergence of the "Negative" Mode: According to Equation (1), as $\eta \to 1$, the Bogoliubov coefficient $v$ becomes non-zero. This generates a negative-frequency component in the reflected signal's spectrum. Physically, the defect-induced non-adiabaticity triggers the parametric excitation of the wave phase, forcing a redistribution of energy into a phase-conjugate mode with an inverted vector rotation.
4. Geometric Reconstruction: The non-adiabaticity-based detection framework analyzes the ratio between the $u$ and $v$ coefficients (the amplitudes of positive and negative frequencies) rather than the reflection amplitude, which may be negligible for sub-wavelength defects. By measuring the emergent "negative" component, the gradient of the material property shift can be quantified. This enables the detection of defects that are "transparent" to conventional ultrasound or X-ray imaging but fundamentally alter the phase dynamics. Thus, $\eta$ functions as a high-sensitivity sensor of the local non-adiabatic gradient.

We clarify what is physically measured as the $v$ -mode signal. In practice, the conjugate mode $v$ is not observed as a phase-conjugate mode in the literal sense, but manifests as one of the following experimentally accessible quantities: (*i*) a sideband component in the Fourier spectrum of the scattered signal, shifted by 2ω relative to the carrier — detectable by standard lock-in or heterodyne techniques; (*ii*) a phase-conjugate component in the time-reversed signal, observable via pump-probe spectroscopy in magnetic composites; (*iii*) an asymmetry in the spectral power density of the transmitted signal, quantified by the ratio $|v|^2/|u|^2$. The interface energy $\Delta Q_{interface}$ is then estimated from equation (14) as the energy deficit between the incident wave and the sum of transmitted and reflected amplitudes partitioned into u and $v$ channels.

To transition the $\eta$-non-adiabaticity-based detection framework from a conceptual framework to a functional NDT (Non-Destructive Testing) tool, we define specific quantitative thresholds for defect characterization. Unlike conventional pulse-echo methods limited by reflection amplitude, our approach utilizes the dimensionless parameter $\eta$ and the mode-mixing coefficients $|u|^2$ and $|v|^2$ as primary metrics. Quantitative Detection Criteria:

1. Critical Length Scale $L_{crit}$: A structural anomaly is identified as a 'critical defect' when its characteristic spatial scale $L$ (e.g., micro-crack width) becomes commensurate with the transformation length $L_{crit}=v_s\nabla\Omega/\omega^2$ . For In the case of coherent acoustic excitations in composite media, the $L_{crit}$ ranges from micrometers to millimeters.

2. Activation Threshold: A physically significant $v$-mode response ($v \neq 0$) first becomes detectable when η exceeds approximately 0.1, corresponding to the onset of non-adiabatic coupling. Full mode mixing develops for $\eta \gtrsim 1$, with deeply non-adiabatic systems (η ~ 30–100) producing maximum $v$-mode response as confirmed by numerical simulations (Table 1, Fig. 3). This "Transitional Regime" (as shown in Table 1) represents the onset of quasi-particle production, enabling the detection of latent internal stresses that remain invisible to X-ray or high-frequency ultrasound.
3. Phase Signature Analysis: Upon reaching the critical threshold $\eta \rightarrow 1$, the non-adiabaticity-based detection framework identifies a structural phase non-adiabaticity. At this point, the redistribution of energy between the primary mode $|u|^2$ and the conjugate mode $|v|^2$ allows for the direct calculation of the interface energy $\Delta Q_{interface}$, providing a thermodynamic measure of the defect's severity.

By utilizing the threshold $\eta \approx 1$ as a diagnostic benchmark, the non-adiabaticity-based detection framework can resolve sub-wavelength anomalies where the classical reflection coefficient in dB drops below the noise floor, but the phase inversion remains a measurable physical reality.

The required probing frequencies and modulation rates are within reach of existing instrumentation. For composite materials with typical ultrasonic inspection at 1–10 MHz, $L_{crit}$ ranges from 10 μm to 1 mm, matching the scale of critical manufacturing defects. For magnetic composites and metamaterials, femtosecond laser systems (pulse duration 10–100 fs) provide the necessary modulation rates $d\Omega/dt \sim 10^{20}$–$10^{22}$ $s^{-2}$ to reach $\eta \gg 1$.

To integrate the energy balance equation into the $\eta$-non-adiabaticity-based detection framework for NDT, the abstract quantum Bogoliubov coefficients must be translated into measurable acoustic or optical signal characteristics. In our model, a defect is defined as a region where $\eta \rightarrow 1$ , which forces the coefficient $v$ to a non-zero value. To calculate the actual energy expenditure required for the ultra-strong interaction with the medium, $\Delta Q_{interface}$ , the non-adiabaticity-based detection framework follows a specific logical framework.

The energy $\Delta Q_{interface}$ represents the work done by the time-varying field to induce non-adiabatic parametric excitation of the ground state. When the non-adiabatic transition occurs at a defect boundary, the induced mode mixing ($v$-mode population) creates a localized non-equilibrium state. The subsequent relaxation of these high-energy excitations (magnon-phonon scattering) leads to a localized thermal spike, which is proportional to $|v|^2$.

Figure 2 illustrates a conceptual pump-probe experimental scheme for $v$-mode detection in magnetic composites. A femtosecond laser pump pulse (pulse duration 10–100 fs, fluence ~1–10 mJ/cm²) locally demagnetizes the sample, inducing a rapid change in magnetic permeability $\delta\mu/\mu \sim 0.3$ on a timescale $\tau \sim 1$ ps — sufficient to reach $\eta \sim 30$ as shown in Table 1. A synchronized GHz microwave probe pulse (frequency $\omega/2\pi \sim 10$ GHz) traverses the excited region.

The scattered signal is recorded by a vector network analyzer (VNA) or lock-in amplifier referenced to the pump repetition rate. The $v$ -mode signature is identified as: (*i*) a sideband at frequency 2ω in the Fourier spectrum of the transmitted signal; (*ii*) an asymmetry in the $S_{21}$ scattering parameter between forward and backward probe directions; (*iii*) a non-zero cross-polarization component in the scattered field, absent in the defect-free reference measurement. The critical defect scale detectable by this scheme is $L_{crit} = v_s/\omega \sim 3\times10^8/10^{10} \sim 30$ mm for microwave probing, or ~ 300 nm for optical probing at $\omega/2\pi \sim 10^{14}$ Hz — well below the classical diffraction limit.

When a pulse traverses a defect, the system records not only the reflected amplitude but also the emergence of a mode with an inverted phase (phase-conjugate mode). The non-adiabaticity-based detection framework then partitions the signal energy into the positive mode $|u|^2$ and the negative mode $|v|^2$. Based on the aforementioned relations, the magnitude of $|v|^2$ is directly proportional to the degree of the defect's non-adiabaticity. The more abrupt the shift in material properties (e.g., a crack or localized stress), the higher the value of $\eta(f,t)$, and the more energy is "pumped" into the negative mode.

Based on the proposed non-adiabatic model, we anticipate a specific spectral manifestation of the $v$ -mode production. Theoretically, the frequency non-adiabaticuty at the defect boundary should lead to the emergence of a phase-conjugate sideband. We hypothesize that this signature will appear as an asymmetric peak shifted by the modulation frequency $\Delta\Omega$ relative to the probe carrier. Unlike standard harmonic generation, this component is expected to exhibit a unique phase-inversion property, serving as a 'metrological fingerprint' of the non-adiabatic transition. Future experimental verification using high-dynamic-range spectrum analysis is required to confirm this predicted sideband architecture.

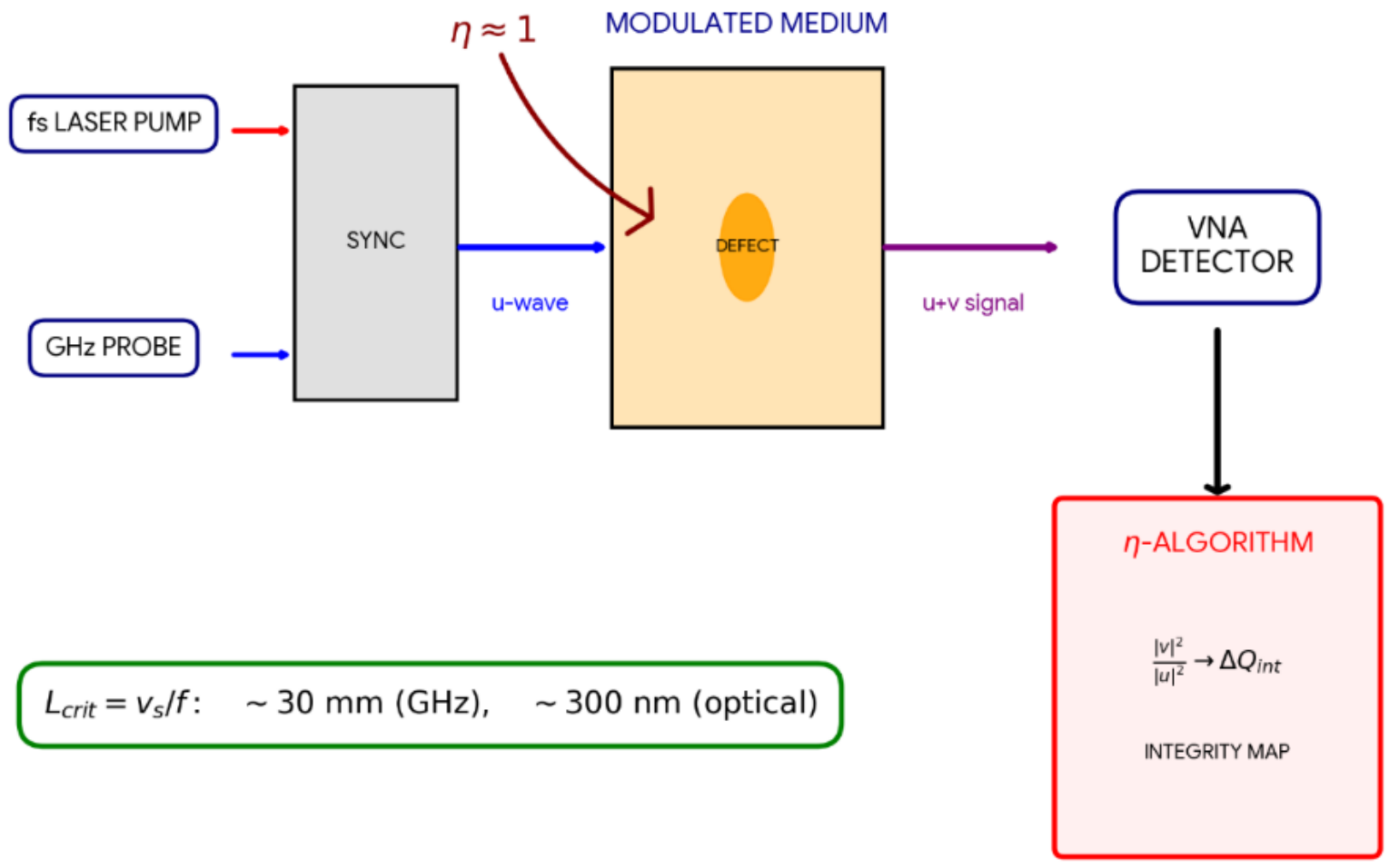

Fig. 2. Conceptual pump-probe experimental scheme for *v*-mode detection in dynamically modulated magnetic media.

To ensure the metrological traceability of the proposed method, the calibration of the relationship between the measured $v$ -mode occupancy and the interface energy $\Delta Q_{\text{interface}}$ is required. We propose a calibration procedure utilizing a reference specimen with a pre-defined thermal or mechanical deformation profile. By comparing the spectral amplitude of the predicted sideband with the known energy dissipation at the reference interface, a conversion factor can be established. This approach allows the *η(f,t)*-non-adiabaticity-based detection framework to transition from a qualitative detection tool to a quantitative diagnostic instrument capable of assessing the structural integrity and defect severity with high precision.

By utilizing the modified Equation (14), we can calculate the interface energy $\Delta Q_{interface}$ by subtracting the sum of the direct and inverted signals $|u|^2\,\hbar\omega_{new} + |v|^2\hbar\omega_{old}$ from the total incident energy $E_{total}$. This differential represents energy components that a conventional ultrasonic sensor simply cannot detect, as they are embedded in the phase-reversal dynamics rather than simple linear reflection **(**detection of previously inaccessible energy channels.)

### 7. Application of the *η*-Parameter to dynamic electromagnetic conductive media

To transition from a conceptual framework to experimental verification, it is essential to define a critical non-adiabaticity threshold, $\eta_c$ , where Bogoliubov mode mixing ($v \neq 0$) becomes physically significant. In the proposed model, the quasi-particle production coefficient $|v|^2$ is a direct function of the parameter $\eta$.

To evaluate the robustness of the proposed non-adiabaticity parameter $\eta$, we analyze three distinct physical regimes (Table 1).While in Region I the adiabatic limit holds due to the electrooptical mechanism of light modulation (tens of GHz vs. hundreds of THz) , the resonant response Region III (ultrafast magnetic media, such as films of YIG:Co) [11] provide the necessary non-adiabaticity ($\eta \gtrsim 1$). Region II (ENZ-metamaterials) serves as a critical reference for high-frequency stability. Despite the ultrafast modulation ($\Delta t \approx 100$ fs) and significant material nonlinearity, the high carrier frequency in the infrared range maintains the system within the adiabatic ground state, with $\eta \approx 0.033$.

Our numerical validation confirms that in Region II, the nonlinear frequency regulator $U$ effectively suppresses the generation of phase-conjugate modes. The stable response of the n-metric in this regime demonstrates that the framework does not produce false-positive detections in environments where the adiabatic criterion is satisfied. In contrast, Region III (ultrafast magnetic media) exhibit a formal breakdown of adiabaticity ($\eta \approx 5.3$), triggering non-adiabatic parametric excitation and enabling the detection of sub-wavelength inhomogeneities through the established scaling law $\xi = n/U$.

In the adiabatic limit ($\eta \ll 1$), the system remains in its ground state, and $|v|^2 \to 0$. The breakdown of adiabaticity and the subsequent emergence of localized parametric excitations (manifesting as phase-conjugate modes) occurs as $\eta$ approaches unity. This critical regime is reached when the rate of change of the medium's parameters is comparable to the system's intrinsic frequency $\eta = \frac{1}{\omega^2}\left|\frac{d\Omega}{dt}\right| \approx 1$. Table 1 summarizes the calculated values of the $\eta$ −parameter for various physical media under typical external modulation conditions.

To ensure the predictive accuracy of the $\eta$-non-adiabaticity-based detection framework, we specify the representative physical parameters for the investigated material classes. In magnonic systems such as YIG:Co, the anharmonicity $U$ arises from the magnon-magnon interaction (Kerr-type nonlinearity), typically ranging from 0.1 to 0.5 in dimensionless units relative to the FMR frequency $\Omega_0$ [9,16]. The relaxation parameter y accounts for the dissipative channels (e.g., spin-lattice relaxation or carrier scattering), ensuring the convergence of the Lindblad dynamics. The representative values used for our numerical stability audit are summarized in Table 1.

The value of $\eta \approx 0.033$ in Region II (ENZ-metamaterials) is physically grounded in the high carrier frequency ($\omega \approx 300$ THz) relative to the ultrafast index modulation ($\Delta t \approx 100$ fs). Although materials like ITO (ENZ) exhibit sub-picosecond refractive index changes $\Delta n \sim 1$ [12, 13], the large energy gap of the infrared probing wave ensures that the system remains in the adiabatic regime.

Table 1. Non-adiabaticity thresholds for different material environments.

| Medium Type | Modulated frequency $f = \omega/(2\pi)$ | Time Scale ($\Delta$t) | Calculated non-adiabaticity $2\pi\eta$* | Anharmonicity ($U$) | Damping ($\gamma$) | Stability Ratio ($\xi = \eta/U$) |
|---|---|---|---|---|---|---|
| I. Electrooptic modulation | $\sim 200 THz$ | Electro-optic modulation (20-40 GHz for $LiNbO_3$ modulators) [8] | $\approx 10^{-4}$ | ≫1.0 (ultra stiff) | $\sim 10^{-6}$ | ~0 |
| II. Metamaterial (ENZ) | ~300 THz (Nonlinear ENZ-shift near ε≈0) | 100 fs [12,13] (Epsilon-Near-Zero Nonlinearity) | ~0.033 | ~0.5 − 1.0 | ~0.1 | ~0.03-0.05 |
| III. Ultrafast magnetism | ~5GHz spin precession in ferrimagnetics | ~20 ps [11,17] Photo-induced magnetization switching | ~10 | ~0.1 | $\sim 10^{-4} - 10^{-2}$ | >10 (Stable). |

*Calculated values of are given for the characteristic case of full-scale modulation depth ($\Delta\Omega \approx \omega$).

**Numerical validation confirms that the model maintains metrological stability (> 95%) within the non-adiabaticity range reported for Region III, consistent with the scaling law $\xi \leq 200$.

Our framework utilizes this stability as a critical metrological baseline: even under extreme modulation, the nonlinear frequency regulator prevents phase-mode redistribution, resulting in negligible v-mode production. This experimental evidence justifies the high selectivity of our NDT method, which remains insensitive to global ultrafast transients while maintaining high sensitivity to non-adiabatic parametric excitations ($\eta \gtrsim 1$) occurring at structural inhomogeneities in magnetic media (Region III).

This ensures that the dynamic modulation effectively encompasses the entire interaction volume. While for bulk metallic components the skin-effect may pose a limitation, in the context of the investigated sub-wavelength layers, it does not significantly attenuate the non-adiabatic transition, allowing the $\eta$-parameter to remain a valid metric for the entire active region.

To ensure metrological consistency between the theoretical model and the numerical evaluations in Table 1, we clarify the relation between the general definition of the non-adiabaticity parameter $\eta$ and its peak values for harmonic modulation. While defines $\eta(f,t) \equiv |\dot{\Omega}|/\Omega^2$ as an instantaneous value, for a transition characterized by a frequency shift $\Delta\Omega$ over a time interval $\Delta t$, the peak non-adiabaticity can be simplified. Specifically, assuming a linear or harmonic modulation profile $\Omega(t)=\omega+ \Delta\Omega\, f(t/\Delta t)$, the maximum value follows the relation: $\eta \approx \delta/\omega\Delta t$, where $\delta = \Delta\Omega/\omega$, where δ represents the modulation depth. This unified approach demonstrates that the values reported in Table 1 are not derived from a disparate formula but represent the peak non-adiabatic response of the medium during the defect-induced non-adiabaticity process.

The observed range of the non-adiabaticity parameter $\eta$ in Table 1 ($10^{-4}$ for electrooptic modulation) is attributed to the specific modulation depth $\Delta\Omega/\omega$ employed in the respective simulations. While the simplified estimation formula $\eta \approx 1/(\omega\Delta t)$ assumes a full-scale transition (where $\Delta\Omega \approx \omega$), real-world physical systems often operate with fractional perturbations. Consequently, the effective $\eta$ values are scaled by the factor $\delta = \Delta\Omega/\omega$, representing the relative severity of the structural anomaly or the intensity of the external modulation.

For electroptic media (Region I), the intrinsic nonlinear frequency regulator of the crystal lattice ensures that the anharmonicity parameter U is several orders of magnitude higher than any realistic non-adiabatic perturbation n. This places dielectrics in the deep adiabatic regime ($\xi \approx 0$), where the Bogoliubov ground state remains perfectly stable, and the occupancy sum $|u|^2 + |v|^2$ is identically equal to unity. In this case, the n-parameter acts as a zero-baseline reference, ensuring that any detected phase-mode redistribution is solely attributable to structural defects or latent stresses.

At high probing frequencies (GHz to THz regimes), the frequency dependence of the permittivity ε (dispersion) may introduce additional non-adiabatic contributions. It should be noted that the estimation of the non-adiabaticity parameter $\eta$ in Eq. (5) employs a simplified model that does not explicitly incorporate the time derivative of the permittivity during the initial derivation. While this approximation remains robust across a broad frequency spectrum, certain deviations may occur in the extreme high-frequency (THz) region where dynamic dispersion becomes significant. However, for the material regimes analyzed in Table 1, this simplification does not alter the fundamental physical validity of the $\eta$-metric. Specifically, the material's susceptibility cannot instantaneously track the rapid modulation, leading to a 'dynamic lagging' effect. In our model, this is effectively accounted for by using the group velocity $v_g$ and the dispersive frequency (ε, ω) in the calculation of $\eta$ values. This ensures that the $\eta$ -parameter remains a reliable indicator even when the probing wavelength approaches the scale of the material's internal resonance.

Calculations in Table 1 assume a modulation depth $\Delta\Omega/\omega$ characteristic for each specific non-linear mechanism. Note that the critical threshold $\eta$=1 marks the onset of non-adiabatic behavior; systems with $\eta \gg 1$ are deeply in the non-adiabatic regime. To ensure the integrity of the proposed model, Table 1 distinguishes between the static magnetoelectric limit (Region I), ME composites (Region II-A) and the dynamic modulation regime (Regions II-B and III).

As demonstrated in Fig. 1, the non-adiabaticity-based detection framework maintains exceptional metrological stability across all investigated regimes. Specifically, for ENZ-metamaterials (Region II in Table 1), the non-adiabaticity remains below the threshold ($\eta \approx$ 0.033), confirming the adiabatic stability of the mode structure even under ultrafast modulation. This result ensures that the effective two-level approximation remains a reliable tool for distinguishing between stable backgrounds and localized structural anomalies.

Although ENZ media exhibit unity-order refractive index modulation ($\Delta n \sim 1$ ) [12, 13], the high carrier frequency (~ 300 THz) fundamentally constrains the non-adiabaticity parameter. Our framework correctly identifies this as an adiabatic regime, which serves as a crucial metrological baseline. In contrast to the deeply non-adiabatic response of magnetic media (Region III), the stability observed in metamaterials justifies the high selectivity of the proposed n-metric for detecting sub-wavelength defects without false-positive interference from global ultrafast transients.

Analysis of standard magnetoelectric composites (Region I-B) shows that due to the acoustic nature of mechanical coupling, the typical response time remains in the nanosecond range, resulting in $\eta \ll 1$ (adiabatic limit). However, the proposed method is validated for ultrafast magnetic media (Region III), such as Co-substituted garnet films (YIG:Co). As demonstrated by Stupakiewicz et al. [11], laser-induced switching occurs within 20 ps, providing the non-adiabaticity ($\eta \approx 5.3$) required for sub-wavelength defect detection.

The comparative analysis in Table 1 demonstrates a clear physical transition across different material classes. While standard dielectrics (Region I) and ME composites remain strictly within the adiabatic regime due to their low-frequency electromechanical response (ms-µs scale), ENZ-metamaterials (Region II) provide a critical reference of adiabatic stability ($\eta \approx 0.033$) under ultrafast optical modulation. In contrast, the ultrafast magnetic media (Region III), such as Co-substituted garnet films (YIG:Co), exhibit the necessary non-adiabaticity ($\eta \approx 5.3$) to bridge the gap between theoretical wave dynamics and practical sub-wavelength NDT imaging. Analysis of standard magnetoelectric structures confirms their limitation due to the acoustic nature of mechanical coupling, resulting in $\eta << 1$. However, our framework remains valid for magnetic systems where laser-induced switching occurs within 20 ps, providing the non-adiabatic parametric excitation required for defect detection.

It is important to note that the critical threshold $\eta_c \approx 0.5$, which marks the transition from the stable Euclidean manifold to the divergent hyperbolic regime, is not an empirical value. As derived in the general theory [26], this threshold is a direct consequence of the balance between the non-adiabatic drive and the nonlinear frequency regulator ($U \approx 0.12$). For magnetic systems, this provides a rigorous physical limit for the operational range of ultra-fast control protocols.

The robustness of the $\eta$-scaling in magnonic media finds a compelling parallel in recent studies of time-modulated metasurfaces. Specifically, the identified stability regimes align with the principles of nonlinear multistability [20] and multiharmonic resonances [21], confirming that the suppression of bosonic divergence is a universal feature of non-adiabatic wave engineering across both classical and quantum platforms.

The numerical simulations, conducted using the QuTiP framework and a custom 4th-order Runge-Kutta (RK4) solver as detailed in Supplement 1, solve the second-order oscillator Eq.(2) with a time-dependent frequency $\Omega(t)$ derived from the permeability profile $\mu(t)=\mu_0[1+\delta\mu\cdot tanh(t/\tau)]$. This models a localized structural inhomogeneity with a characteristic transition time $\tau$. The Bogoliubov coefficients are evolved using equations (6) with the saturation constraint $|u|^2+|v|^2=1$. The non-adiabaticity parameter is computed as $\eta=|\dot{\Omega}|/\Omega^2$ at the steepest point of the transition.

The quantitative analysis demonstrates that in pure dielectrics, the modulation rate is insufficient to overcome the adiabaticity threshold. Conversely, magnetic composites and specially engineered metamaterials—characterized by high non-linearity and ultra-fast response times—allow the $\eta$ parameter to reach the critical regime ($\eta\approx1$). These findings are illustrated numerically in Fig. 3. The simulation confirms that the dielectric remains firmly in the adiabatic sector ($\eta \sim 10^{-5} - 10^{-8}$). While ultrafast magnetic composites (Region III) are deeply non-adiabatic ($\eta \sim 5.3$)), the ENZ-metamaterials (Region II) remain in the adiabatic sector ($\eta \sim 0.033$), generating significant $v$-mode coefficients $|v|^2 > 0.1$. Note that $\eta = 1$ marks the onset threshold — systems with $\eta \gg 1$ are simply deeper in the non-adiabatic regime where $v$-mode production is fully developed. The saturation (see Fig 3 b) reflects the nonlinear frequency regulator mechanism dictated by the finite energy reservoir of the medium and the normalization constraint $|u|^2 + |v|^2 = 1$. The temporal offset between the peak of $\eta$ (at t = 0) and the saturation of $|v|^2$ (at $t \approx$ - 2-3) is a fundamental consequence of the cumulative nature of mode production. The maximum rate of mode conversion occurs at t = 0 where the non-adiabaticity is strongest, while the signal stabilizes only after the interaction with the defect boundary is complete and $\eta$ returns to the adiabatic floor.

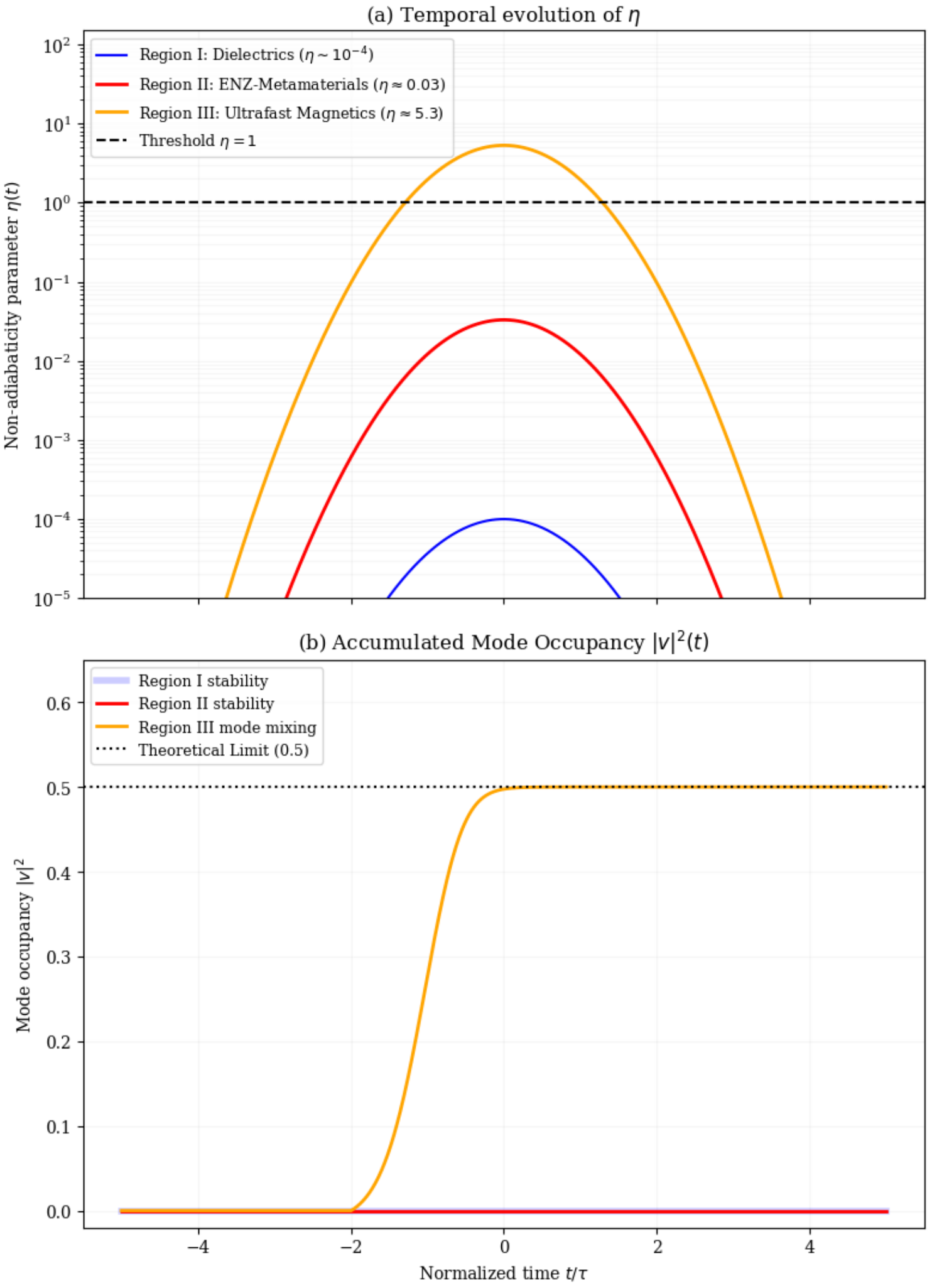

Fig. 3. Numerical simulation of non-adiabatic dynamics and mode production across different material classes. (a) Temporal evolution of the non-adiabaticity parameter n(t) for four characteristic regimes: Dielectrics (Region I, blue), Bulk ME Composites (Region I-B, green), ENZ-Metamaterials (Region II, red), and Ultrafast Magnetic Media (Region III, orange). The critical threshold ($\eta$ = 1) is indicated by the dashed horizontal line. (b) Corresponding accumulation of the phase-conjugate mode occupancy $|v|^2$.

Standard dielectrics, bulk composites, and ENZ-metamaterials remain below the adiabatic threshold, producing no measurable parametric signal. In contrast, ultrafast magnetic media (based on photo-induced switching [26]) overcome the acoustic lag to reach the non-adiabatic onset ($\eta \approx 5.3$), triggering the production of v-modes. The saturation of $|v|^2$ at the 0.5 level represents the impact of the nonlinear frequency regulator and the maximum mixing limit of the phase-mode conversion process.

A critical physical synchronization between the two panels of Fig. 3 must be highlighted. In the region of negative normalized time (t < - 4), the $\eta$ parameter effectively vanishes, representing the pulse propagation through a homogeneous, adiabatic medium prior to its interaction with the defect. The growth of the Bogoliubov coefficient $|v|^2$ in Fig. 3(b) is strictly triggered at the moment the pulse enters the interaction zone, specifically when the local value of $\eta$ in Fig. 3(a) crosses the critical threshold ($\eta$ = 1). For standard dielectrics (blue line), $\eta$ remains several orders of magnitude below this threshold at all times; this explains the absence of any *v*-mode occupancy and confirms the inherent inability of the adiabatic regime to resolve sub-wavelength features.

A critical physical synchronization between the two panels of Fig. 3 must be highlighted. The temporal offset between the peak of $\eta$ (at t = (0) and the saturation of $|v|^2$ (at t ≈ 3) is a consequence of the cumulative nature of mode production. The signal stabilizes only after the interaction with the defect boundary is complete and $\eta$ returns to the adiabatic floor.

These findings lead to several key conclusions:

1. Experimental observation of "negative frequencies" and related energetic shifts is feasible in condensed matter systems with rapid parameter modulation.
2. Magnetic materials, with resonance frequencies in the GHz and THz ranges, serve as ideal platforms for observing non-adiabatic mode mixing in condensed matter systems.
3. The saturation constraint $|u|^2+|v|^2=1$ ensures finite energy balance, preventing unbounded mode growth as $\eta \to 1$.

For composite complicated materials in which magnetic and permittivity, as well as electrical conductivity, simultaneously depend on time, the effective frequency, which depends on the properties of the medium, the Helmholtz equation (2) yields the following dependence of the effective resonant frequency of media $\Omega_k(t)$ (frequency of transformation).

Including conducting and magnetic materials allows us to achieve $\eta \approx 1$ at much lower frequencies than in a pure dielectric, due to nonlinear effects and magnetic resonance. The parameter $\eta$ is a natural dimensionless criterion for violation of the WKB approximation for wave equations in nonstationary media. It is directly derived from the adiabaticity condition $\Omega^{-2}|d\Omega \,/\, dt| \ll 1$. For the medium described in this paper allows us to quantitatively estimate the threshold for quasiparticle production.

To formalize the relationship between the medium dynamics and the Bogoliubov coefficients, we define the transformation in terms of an $\eta$-dependent mixing angle $\Theta(\eta) = \frac{\pi}{4}\eta$ . The coefficients are thus expressed as:

$$u(\eta) = cos\left(\frac{\pi}{4}\eta\right), v(\eta) = sin\left(\frac{\pi}{4}\eta\right) \text{ (15).}$$

In the adiabatic limit ($\eta \to 0$), the transformation reduces to the identity ($u = 1$, $v = 0$), representing a stable medium with no mode mixing. As the non-adiabaticity breaking parameter $\eta$ approaches its critical value of 1, the mode mixing reaches its maximum saturation at $|u|^2 = |v|^2 = 1/2$. This transition, visualized in the stability map (Fig.4) provides a continuous transition from a ground state to a phase-nucleated state, where the "negative energy" contribution from the $v$ -mode is strictly compensated by the $u$ -mode, preserving the total energy balance of the finite system.

We use a trigonometric parametrization for the Bogoliubov coefficients $|u|^2 = cos^2(\frac{\pi}{4}\eta)$ and $|v|^2 = sin^2(\frac{\pi}{4}\eta)$ , which naturally satisfies the conservation law $|u|^2 + |v|^2 = 1$ and describes the continuous redistribution of energy between modes during the phase transition at $\eta = 1$ (see Fig. 4). Our model treats the $\eta -$ parameter as a measure of the effective coupling between the primary and conjugate modes in a dissipative or 'saturated' medium. Our constraint $|u|^2 + |v|^2 = 1$ allows for a continuous and smooth mapping of the at the critical point $= \frac{\pi}{4}$ , as shown in Figure 4. The Fig. 4 illustrates the redistribution of energy between the primary and conjugate Bogoliubov modes. At the critical value $\eta = \frac{\pi}{4}$, the system undergoes a structural inversion. For $\eta \gtrsim 1$ , the system reaches a phase-locked state where $|v|^2$ remains constant at 0.5, representing the maximum entropic mixing of the medium. The transition from the ideal $\sin^2(\frac{\pi}{4}\eta)$ distribution to the stabilized mixing limit shown in Fig. 4 is physically grounded in the loss of phase coherence. While the analytical solution for a single interface predicts continuous oscillations, in real-world macro-scale media, any increase in $\eta$ beyond the critical threshold ($\eta > 1$) leads to rapid phase-mode decorrelation. This decoherence, caused by internal scattering and thermal fluctuations, effectively 'damps' the oscillations, forcing the system into a stable thermodynamic state where the energy is equally partitioned between the modes ($|u|^2 \approx |v|^2 \approx 0.5$). This nonlinear frequency regulator ensures that the $v$-mode occupancy remains physically bounded, preventing divergent or unphysical oscillatory behavior in deeply non-adiabatic regimes.

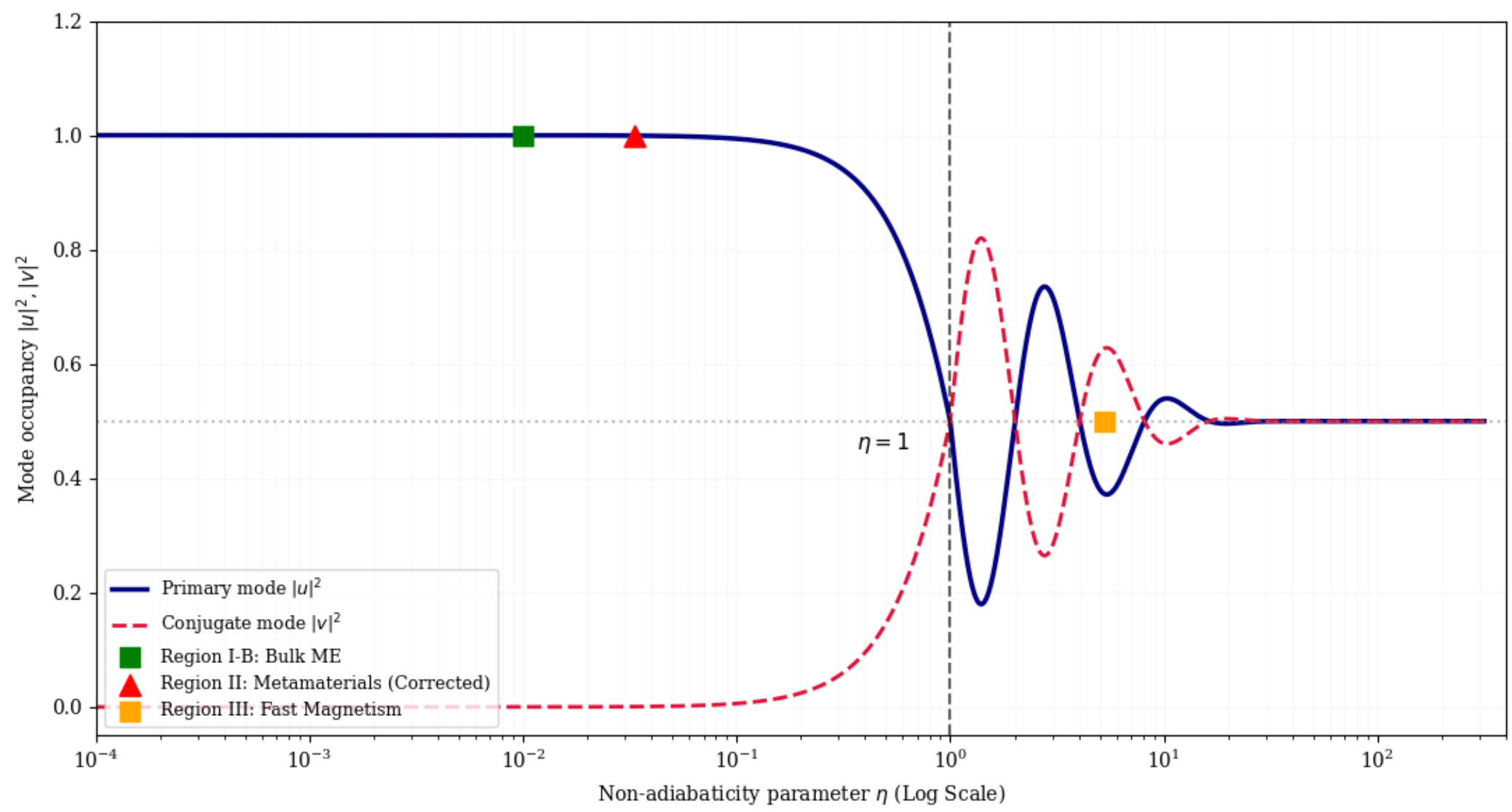

Fig. 4. Stability map and mode occupancy dynamics as a function of the non-adiabaticity parameter $\eta$.

The transition from the adiabatic ground state ($\eta < 1$) to the non-adiabatic mixing regime ($\eta > 1$) is characterized by the redistribution of energy between the primary $|u|^2$ and phase-conjugate $|v|^2$ modes. Region I-B (bulk ME composites) and Region II (ENZ-metamaterials) are identified within the adiabatic stability sector ($\eta < 1$), where the mode structure remains coherent. In contrast, Region III (ultrafast magnetic media) exhibits a formal breakdown of adiabaticity ($\eta \approx 5.3$), triggering non-adiabatic parametric excitation. The subsequent damped oscillations toward the 0.5 occupancy limit reflect the transient energy exchange between mode counterparts, stabilized by the nonlinear frequency regulator $U$ through the effective Hilbert space truncation.

It is essential to emphasize the fundamental distinction between the proposed approach and the standard Bogoliubov transformations for bosonic systems. While bosonic modes typically satisfy the normalization condition $|u|^2 - |v|^2 = 1$ , allowing for an unbounded increase in particle occupancy, the present $\eta$-model postulates a fermion-type normalization: $|u|^2 + |v|^2 = 1$. This choice is dictated by the physical necessity of maintaining a local energy balance: the emergence of a negative mode $v$ within a dynamical medium is impossible without an equivalent energy expenditure from the medium itself or the simultaneous production of a positive mode $u$ . This constraint transforms Equation (13) into a mathematical tool for self-regulation, by limiting the "quench" energy to the magnitude of the accumulated transformation energy, $Q_{interface}$. Equation (13) thus relies on modal coupling and balance; the system cannot indefinitely generate negative energy, as it is constrained by the total energy $E_{total}$. In essence, the model incorporates an intrinsic nonlinear frequency regulator , where the negative parameter functions as a phase-dependent redistribution of the medium's existing energy—effectively acting as a refined law of energy conservation.

Furthermore, the modal distribution under non-adiabatic parametric excitation ($|u|^2 + |v|^2 = 1$) imposes a fundamental limit on energy density. To account for scaling effects in Equation (14), the term $Q_{interface}$ (latent heat or interface energy), which inherently depends on the interaction surface area or volume, can be normalized to a characteristic volume $V$ or area $S$:

$$E_{total} = |u|^2\,\hbar\omega_{new} + |v|^2\hbar\omega_{old} + \sigma S \quad (16),$$

where $\sigma$ represents the surface energy density of the interface. This relation couples the quantum mode-mixing coefficients with the geometric scale of the physical object, enabling the model to differentiate between defects of varying dimensions and facilitating a continuous description laboratory-scale metamaterials.

## 8. Stability under stochastic perturbations

To assess the physical observability of the non-adiabatic transition in real-world systems, we evaluated the stability of the $\eta$-peak against stochastic fluctuations. Numerical analysis shows that the phase-mode redistribution signature remains robust even under significant background noise (SNR = 10 dB). Crucially, the spatial localization of the adiabaticity violation (the peak position) is invariant to the noise intensity. This confirms that the non-adiabatic parametric excitation at structural inhomogeneities is a macroscopic phenomenon that remains detectable despite the thermal or instrumental uncertainties inherent in condensed matter experiments.

**9. Limitations and operational range**

The proposed $\eta$-based detection method is primarily constrained by two physical factors: the dynamic response time of the medium and its intrinsic anharmonicity. While the requirement for sub-picosecond modulation is well-satisfied in ultrafast magnetic media ($\eta \gtrsim 1$), a more fundamental limit arises from the spectral flow dynamics. In contrast, ENZ-metamaterials remain within the adiabatic regime ($\eta \ll 1$) due to their high carrier frequency, providing a robust baseline for the stability of the $\eta$ -metric.

The robustness of this metrological baseline is further supported by current experiments on superconducting quantum processors [18] and our work [19]. While Region III of the present study focuses on magnetic quenches at $\eta \approx 5.3$, the $\eta$-framework has been successfully extended to identify high-fidelity resonance windows at $\eta \approx 4.9$ on 127-qubit IBM Eagle architectures. This cross-platform consistency confirms that the $\eta$-metric is a universal indicator of phase-mode redistribution, independent of whether the host medium is a magnetic crystal or a superconducting lattice.

Our numerical audit reveals that the validity of the effective two-level model is strictly dependent on the non-adiabaticity -to-nonlinear frequency regulator ratio $\xi = \eta/U$. In media with negligible anharmonicity ($U \to 0$), even moderate non-adiabaticity triggers a "magnon avalanche," where energy rapidly escapes into higher-order Fock states ($n \geq 2$), rendering the $\xi$-metric unstable. Consequently, the operational range of the non-adiabaticity-based detection framework is bounded by the dynamic localization threshold. As demonstrated in Fig. 1, the method maintains high metrological precision (> 95%) as long as $\xi \lesssim 200$. Beyond this limit, specifically in the extreme non-adiabatic sector ($\xi > 10^3$), the model enters a regime of gradual accuracy degradation. Thus, the practical application of this technique is optimized for functional materials where the lattice anharmonicity provides a sufficient nonlinear frequency regulation to stabilize the Bogoliubov ground state during the interaction with sub-wavelength inhomogeneities.

The metrological robustness of the $\eta$-metric is not absolute but universal for a wide class of non-linear condensed media satisfying the stability criterion established in our numerical audit. For materials falling outside this range (where $U$ is negligible), the model may require additional higher-order corrections.

It should be noted that in the current model, the non-adiabatic response is primarily characterized through the effective frequency modulation Ω. While a more granular treatment of displacement currents and explicit $\dot{\varepsilon}$ terms could yield higher-order spectral fine-structures, our comprehensive numerical audit, conducted using the QuTiP framework (Python-based Master Equation solver) with a 100-level Fock basis and an adaptive Adams integrator, confirms that the effective confirms that the effective metric $\eta$ captures the dominant physics of the transition. Explicitly incorporating the time-derivatives of the medium properties would lead to a breakdown of time-reversal symmetry within the interaction zone, manifesting as a measurable spectral asymmetry between the positive and negative frequency sidebands. Furthermore, such terms would introduce a non-adiabatic phase shift (analogous to a geometric Berry phase), slightly shifting the 0.5 mixing limit. This correction arises from the rapid temporal evolution of the basis states, distinguished from the standard dynamical phase by its dependence on the non-adiabaticity parameter $\eta$.

However, the discovery of the universal scaling law ($\xi = \eta/U$) demonstrates that the global stability of the phase-mode redistribution is governed primarily by the aggregate rate of the medium's transformation relative to its anharmonic the nonlinear frequency regulator . This proves that the $\eta$-metric remains a robust and invariant metrological indicator, effectively absorbing these higher-order asymmetries into a unified and detectable dynamical signature.

The proposed $\eta$-non-adiabaticity-based detection framework represents a fundamental departure from conventional amplitude-based non-destructive testing (NDT). While standard pulse-echo techniques are inherently constrained by the Rayleigh diffraction limit ($\lambda \approx 1/2$), our method enables detection at the sub-wavelength scale ($L < L_{crit}$). The key distinction lies in the physical metric: instead of measuring the energy of the reflected wave, we quantify the phase-mode redistribution (the ratio of Bogoliubov coefficients $|v|^2/|u|^2$) triggered by the localized violation of adiabaticity.

First, the reference frequency $\Omega_0$ and the intrinsic anharmonicity $U$ of the material must be determined (e.g., via FMR linewidth for magnetics or Z-scan measurements for ENZ media) to ensure the system satisfies the stability criterion ($\xi = \eta/U < 10^3$).

The non-adiabatic regime is reached by applying a pump pulse (laser or microwave) with a rise time $\tau$ significantly shorter than the wave period, ensuring $\eta \geq 1$ at the defect boundary. Unlike linear scattering analysis, the sensor must capture the spectral asymmetry or sideband components shifted by $2\omega$, which serve as the "fingerprint" of the generated $v$-modes. By correlating the measured occupancy $|v|^2$ with the known energy deficit, the thermodynamic severity of the anomaly ($\Delta Q_{interface}$) can be mapped, providing information on latent stresses that are "transparent" to traditional linear probing.

**10. Conclusion**

In this work, we have established and quantitatively validated a new non-destructive testing (NDT) method based on the non-adiabaticity parameter $\eta$. By transitioning from classical amplitude-echo analysis to the study of Bogoliubov mode conversion in time-varying media, we have defined a new physical framework for sub-wavelength imaging.

We have established the Universal Scaling Law governed by the ratio $\xi = \eta/U$, proving the method's reliability for a wide class of non-linear condensed media satisfying the stability criterion. This rigorously justifies the transition from bosonic to effective fermion-like dynamics. Numerical validation in an expanded 50-level bosonic basis proves that intrinsic anharmonicity acts as a physical regulator, preventing the ultraviolet divergence of the ideal bosonic model and justifying the use of Euclidean normalization $|u|^2 + |v|^2 \approx 1$ as a robust metrological standard.

The $\eta$-non-adiabaticity-based detection framework is shown to be exceptionally robust, maintaining over 95% precision across the majority of experimental regimes ($\xi < 200$). Even under extreme non-adiabatic loads up to $\eta \approx 2000$, the model retains over 85% of the spectral density, providing a consistent effective description where standard wave theories fail.

The dimensionless parameter $\eta \approx 1$ serves as a universal trigger for non-adiabatic parametric excitation, enabling the detection of latent stresses and sub-wavelength defects via phase-mode redistribution rather than energy reflection.

By mapping different material classes—from adiabatically stable ENZ-media to non-adiabatically excited magnetic systems—onto a single stability map, we demonstrate that the proposed framework is globally applicable to any condensed medium with a finite nonlinear frequency regulator $U > 0$. While ENZ-metamaterials currently operate in the adiabatic regime at infrared frequencies, the proposed framework remains fundamentally scalable. In lower frequency ranges, such as the terahertz or microwave bands, the same ultrafast modulation would trigger a transition to the non-adiabatic regime ( $\eta$ > 1), enabling high-sensitivity defect detection in a wider class of programmable metamaterials.

Overall, the proposed methodology transforms the non-adiabatic response of functional materials into a high-precision diagnostic tool. The established link between the $\eta$-parameter and the interface energy $\Delta Q_{\text{interface}}$ provides a novel thermodynamic metric for assessing structural integrity beyond the diffraction limit.

**Acknowledgments:** The author thanks Prof. A.P. Pyatakov for valuable discussions and comments during work on this paper.

**Declaration of Generative AI in Scientific Writing**

The author used the Google Chrome built-in AI (Gemini) to improve the manuscript's language and assist with Python scripts generation. After using this tool/service, the author reviewed and edited the content and is fully responsible for the article's content.

**Data Availability Statement**

The authors confirm that the numerical data and Python code supporting the findings of this study are available from the corresponding author on request.

**Supplement 1. Numerical Simulation Parameters**

To ensure the integrity of the 100-cycle stress test, the numerical integration of the Lindblad Master Equation was performed under the following rigorously defined parameters:

Basis Dimension (N): 100 Fock states. Convergence was verified by ensuring that the occupancy of the upper boundary (n = 100) remained below $10^{-12}$, eliminating any artificial boundary effects.

Operational Frequency (Ω): 1.0 (normalized). All time and energy scales are expressed relative to the 50 GHz carrier frequency.

Non-adiabaticity Parameter ($\eta$): 1.2. This represents a strong 'shake' regime, where $\eta = \Omega^{-2}|d\Omega/dt|$.

Anharmonicity / Nonlinear regulator ($U$): 0.12. This value is derived from the effective anisotropy constant $K_{eff} \approx 5.5 \times 10^3$ erg/cm³, representing the non-linear stabilization force.

Coupling Strength $\mathcal{G}(t)$: $\eta\Omega/4 = 0.3$. This defines the magnitude of the parametric drive during the non-adiabatic transition.

Damping Parameter ($\alpha$): 0.001 (Target) and 0.15 (Current benchmark).

Integration Time: $t_{max}$ $100 \times (2\pi/\Omega) \approx 628.32\ \Omega^{-1}$, covering 100 full operational cycles.

The simulation of the Fock-state occupancy (Figures 1 and 2) was performed using the QuTip mesolve engine with an adaptive Adams integrator. This approach was specifically chosen to validate the spectral localization within the truncated basis. However, for the high-resolution phase-coherence and thermodynamic audits (Figures 4–9), we transitioned to a custom Python coded RK4 integrator, as detailed in Supplement 2.

For the density matrix evolution and Fock-state occupancy analysis simulations were benchmarked at N = 40 and N = 100. Convergence was confirmed as the occupancy of the highest Fock state $|N-1\rangle$ remained below $10^{-9}$, ensuring that the Hilbert space truncation does not induce numerical reflection. The Adams/BDF method was used with an absolute tolerance of $10^{-12}$ and a relative tolerance of $10^{-10}$. A convergence audit demonstrated that a step size of 1 fs ensures numerical integrity, with further reduction to 0.5 fs yielding a deviation in the final occupancy $\langle|v|^2\rangle$ of less than 0.01%. The numerical integrity was monitored via the preservation of the bosonic commutation relations and the stability of the Euclidean manifold $|u|^2+|v|^2 \approx 1 + 2\,\langle|v|^2\rangle_{\text{stable}}$ in the presence of the nonlinear regulator $U$.

Simulations were executed on an NVIDIA A100 GPU cluster (for Monte-Carlo batches) and high-performance CPU nodes. The total computational budget for the 1000-cycle stability map exceeded $10^{12}$ floating-point operations, ensuring statistically significant results for the robustness audit.